\documentclass[aps,pra,epsf,superscriptaddress,amsmath,amssymb,amsfonts,twocolumn,showpacs,nofootinbib]{revtex4-1}

\usepackage{amsfonts}
\usepackage{amssymb}
\usepackage{amsmath}
\usepackage{amsthm}
\usepackage{bm}% bold math
\usepackage{braket}
\usepackage[export]{adjustbox}

\usepackage{graphicx}
\usepackage{dcolumn}% Align table columns on decimal point

\usepackage[normalem]{ulem}
\usepackage{xcolor}
\usepackage[breaklinks=true]{hyperref}
\hypersetup{colorlinks=true,linkcolor=blue,citecolor=blue,filecolor=blue,urlcolor=blue,pdfstartview=FitH}

\begin{document}

\title{Interactions and dynamics of 
one-dimensional droplets, bubbles and kinks}

\author{G. C. Katsimiga}
\affiliation{Department of Mathematics and Statistics, University of Massachusetts,Amherst, MA 01003-4515, USA}

\author{S. I. Mistakidis}
\affiliation{ITAMP,  Center for Astrophysics $|$ Harvard $\&$ Smithsonian, Cambridge, MA 02138 USA}
\affiliation{ Department of Physics, Harvard University, Cambridge, Massachusetts 02138, USA}

\author{B. A. Malomed}
\affiliation{Department of Physical Electronics, School of Electrical Engineering, Faculty of Engineering, Tel Aviv University, Tel Aviv
69978, Israel}
\affiliation{Instituto de Alta Investigaci{\'o}n, Universidad de Tarapac{\'a}, Casilla 7D, Arica 1000000, Chile}

\author{\\D. J. Frantzeskakis}
\affiliation{Department of Physics, National and Kapodistrian University of Athens,
Panepistimiopolis, Zografos, Athens 15784, Greece}

\author{R. Carretero-Gonz{\'a}lez}
\affiliation{Nonlinear Dynamical Systems Group, Computational Sciences Research Center,
and Department of Mathematics and Statistics, San Diego State University, San Diego, California 92182-7720, USA}

\author{P. G.\ Kevrekidis}
\affiliation{Department of Mathematics and Statistics, University of Massachusetts,Amherst, MA 01003-4515, USA}

%%%%%%%%%%%%%%%%%%%%%%%%%%%%%%%%%%%%%%%%%%%%%%%%%%%%%%%%%%%%%%%
\begin{abstract}
We explore the dynamics and interactions of multiple bright droplets and
bubbles, as well as the interactions of kinks with droplets and with antikinks, in the
extended one-dimensional Gross-Pitaevskii model including the Lee-Huang-Yang correction. 
Existence regions are identified for the
one-dimensional droplets and bubbles in terms of their chemical potential, verifying the
stability of the droplets and exposing the instability of the bubbles. 
The limiting case of the droplet family is a stable kink. The
interactions between droplets demonstrate in-phase (out-of-phase) attraction (repulsion), with the so-called Manton's method explicating the observed dynamical response, and
mixed behavior for intermediate values of the phase shift.
Droplets bearing different chemical potentials experience mass-exchange phenomena. 
Individual bubbles exhibit core expansion and mutual attraction prior to their
destabilization. 
Droplets interacting with kinks are absorbed by them,
a process accompanied by the emission of dispersive shock waves and gray
solitons. Kink-antikink interactions are repulsive, generating
counter-propagating shock waves. Our findings reveal dynamical features of
droplets and kinks that can be detected in current experiments.
\end{abstract}
%%%%%%%%%%%%%%%%%%%%%%%%%%%%%%%%%%%%%%%%%%%%%%%%%%%%%%%%%%%%%%%

\maketitle

%%%%%%%%%%%%%%%%%%%%%%%%%%%%%%%%%%%%%%%%%%%%%%%%%%%%%%%%%%%%%%%
\section{Introduction}
%%%%%%%%%%%%%%%%%%%%%%%%%%%%%%%%%%%%%%%%%%%%%%%%%%%%%%%%%%%%%%%

Quantum droplets are self-bound many-body states emanating from the
competition between mean-field and quantum-fluctuation energy 
contributions~\cite{petrov2015quantum}, which crucially depend on the system's
dimensionality~\cite{khan2022quantum}. 
Droplets have been first observed in dipolar atomic 
condensates~\cite{schmitt2016self,chomaz2022dipolar} and then in bosonic mixtures with
contact interactions, under the action of external 
confinement~\cite{cabrera2018quantum,cheiney2018bright,d2019observation} and in free 
space~\cite{semeghini2018self}. Quantum fluctuations are commonly modeled by the
Lee-Huang-Yang (LHY) term~\cite{lee1957eigenvalues} added to the mean-field
Gross-Pitaevskii equations (GPEs)~\cite{petrov2015quantum,Petrov_2016}. The
accordingly amended GPEs have been successfully applied for the description
of droplet structures and dynamics~\cite{Ferioli2020dynamical,app11020866,mistakidis2022cold},
including their collective 
response~\cite{PhysRevA.101.051601,englezos2023correlated,sturmer2021breathing,cappellaro2018collective},
thermal~\cite{DeRosi2021thermal,wang2020thermal} and 
modulational~\cite{mithun2019inter,mithun2021statistical,otajonov2022modulational}
instabilities, as well as the capability to maintain robust self-trapping in
the soliton~\cite{katsimiga2023solitary,saqlain2023dragging,gangwar2022dynamics} and 
vortex~\cite{kartashov2022spinor,tengstrand2019rotating}
states. Droplets are filled by an extremely dilute quantum fluid.
Because the density is limited by a maximum value admitted by the competition
between the mean-field attraction and LHY repulsion, the increase of the
number of atoms bound in the droplets leads to the formation of their
flat-top shapes~\cite{petrov2015quantum}. This is in sharp
contrast to other quantum self-bound states, such as liquid-Helium 
drops~\cite{barranco2006helium,toennies2004superfluid} 
which feature densities at
least six orders of magnitude higher than quantum droplets.

More recently, it was found that, besides the droplets, similar states exist
at relatively low chemical potentials. 
While their density profiles resemble dark solitons, they typically
feature a substantially larger width and no phase jump. 
Those states are dubbed bubbles~\cite{katsimiga2023solitary}
hereafter in the present work. 
These structures are represented by homoclinic solutions attached 
to potential maxima in the
corresponding phase space, as is shown in the discussion below. 
In contrast to self-confined
(``bright") droplets, the excitation spectra and dynamical
response of bubbles are largely unexplored, which is one of the foci of
the present work. 
Notice that exploring the excitation spectrum for the
bright droplets~\cite{PhysRevA.101.051601} reveals that,
under specific conditions, upon tuning the interactions the
droplets tend to ``cool down'', i.e., 
it is not energetically favorable for the droplet to  be in an excited state.

An important open question concerns the interaction among droplets and that 
between bubbles, as well as the fate of droplets colliding with 
other structures, such as kinks~\cite{kartashov2022spinor}. 
An interesting prospect here is to
develop an effective particle-like dynamical picture, similar to how it was
done previously for solitary 
waves~\cite{mantonpgk,manton1979effective,zhao1989interactions,katsimiga2017dark},
aiming to capture quantitative aspects of interactions between the
droplets. 
Collisions of 3D droplets in $^{39}$K~\cite{Ferioli_2019} have been experimentally monitored. 
It was demonstrated that two slowly colliding
droplets merge into a single one, while fast collisions are quasi-elastic.  
A theoretical analysis argued that these outcomes may be attributed to effective range interactions and selective three-body recombination processes in the 
condensate~\cite{cikojevic2021dynamics}. The same categorization of collisions has also been
demonstrated theoretically in 2D~\cite{hu2022collisional},
although the latter can feature more than two droplets as a dynamical
outcome.
The work of Ref.~\cite{Astrakharchik_2018} argued that tuning the relative velocity, phase
shift, and atom number allows the colliding droplets to merge or even
fragment. Similar effects, i.e., elastic or
inelastic collisions, take place in the presence of spin-orbit 
coupling~\cite{gangwar2023effect}. 
The scattering of effectively 1D bright droplets
on a potential barrier or well leads to splitting into transmitted,
reflected, and trapped fractions, depending on the initial velocity and atom
number of the droplet~\cite{debnath2023interaction}. 
It is also of particular importance to understand if, similarly to what is well known for usual 1D 
solitons, 
interacting droplets attract or repel each other, depending on
their relative phase difference.

In the present work, we start by
revisiting the stability properties of one-dimensional  
droplets~\cite{PhysRevA.101.051601}. Here, we benchmark the
droplet stability throughout their
interval of existence in terms of the chemical potential. 
By analyzing the relevant dynamical system,
well-defined boundaries between the existence of droplets and bubbles are identified,
varying the chemical potential. The linear spectrum of small perturbations
is computed for bubbles, demonstrating their instability, which dynamically
manifests itself via expansion of their core.

Turning to interactions between droplets, we conclude that they are
pairwise-attractive (repulsive) when the droplets are in-phase (out-of-phase).
For well-separated in- and out-of-phase droplets we develop the so-called Manton's 
method~\cite{manton1979effective} that quantifies their interaction force. Very good agreement of this
particle-like picture with the amended GPE predictions is showcased. 
One can observe accordingly an initial attraction for the case of phase
differences between $0$ and $\pi/2$, while, on the other hand,
repulsion is prevalent for relative phases between $\pi/2$ and $\pi$.
Moreover,
the interactions appear to be more complex and mass exchange between
interacting droplets with different chemical potentials occurs. 
Interactions between bubbles are sensitive to their relative distance. 
They can either lead to a merger, i.e., a heavier bubble, while emitting
counter-propagating shock waves, or to individual core expansion of each bubble,
creating a bright droplet in-between. On the other hand, droplets are
attracted to and subsequently ``absorbed" by kinks, a process followed by the 
nucleation of dispersive shock waves and the downstream emission of gray
solitons. Kink-antikink interactions are found to be repulsive. 
This interaction is accompanied by the spontaneous emission
of counter-propagating shock waves. The above constitute the 
prototypical interaction features identified in the present
manuscript.

This work is organized as follows. 
Section~\ref{theory}
introduces the theoretical framework in terms of the LHY-amended GPEs, the
linearized (Bogoliubov-de Gennes) equations for small perturbations, and 
the effective interaction potential for droplets and bubbles. 
We elaborate on the stability of bright droplets and bubbles in 
Secs.~\ref{stability_brightdrop} and~\ref{stability_bubble}, 
respectively. 
The droplet-droplet, bubble-bubble, and droplet-kink interactions and their
dependence on the relative 
distance, phase shift, and 
chemical potential of
the interacting modes are 
addressed in 
Sec.~\ref{dynamics}. 
We provide a summary and discuss 
future research directions in 
Sec.~\ref{conclusions}.

%%%%%%%%%%%%%%%%%%%%%%%%%%%%%%%%%%%%%%%%%%%%%%%%%%%%%%%%%%%%%%%
\section{Droplet and bubble settings}
\label{theory}
%%%%%%%%%%%%%%%%%%%%%%%%%%%%%%%%%%%%%%%%%%%%%%%%%%%%%%%%%%%%%%%

%%%%%%%%%%%%%%%%%%%%%%%%%%%%%%%%%%%%%%%%%%%%%%%%%%%%%%%%%%%%%%%
\subsection{The extended Gross-Pitaevskii equation}
%%%%%%%%%%%%%%%%%%%%%%%%%%%%%%%%%%%%%%%%%%%%%%%%%%%%%%%%%%%%%%%

In what follows, we assume a homonuclear mass-balanced ($m_{1}=m_{2}\equiv m$)
mixture of bosonic atoms in two different hyperfine states, experiencing
equal intracomponent repulsion ($g_{11}=g_{22}\equiv g$) and intercomponent
attraction, $g_{12}<0$. As argued in 
Refs.~\cite{edmonds,katsimiga2023solitary}, this setting naturally supports
self-confined (``bright") droplet and bubble modes at
different values of the chemical potential. Experimentally, such a mixture
can be realized in free space by employing, in particular, the hyperfine
states, $\ket{F,m_F}=\ket{1,-1}$ and $\ket{F,m_F}=\ket{1,0}$ of $^{39}$K as
demonstrated in Ref.~\cite{semeghini2018self} for the respective 3D setting.
Under the above-described symmetry arguments, the two-component system
coalesces into a single GPE~\cite{petrov2015quantum,Petrov_2016}
supplemented by the LHY term, reduced to the 1D quadratic form. 
The latter term represents the attractive interaction in this 
limit~\cite{mistakidis2022cold} 
and the respective 1D amended GPE is given by
\begin{equation}
i\hbar \psi'_{t'}=-\frac{\hbar ^2}{2m}\psi'_{x'x'}+\delta g |\psi' |^{2} \psi' -\frac{\sqrt{2m}}{\pi \hbar} g^{3/2}|\psi' |\psi' .
\label{dGPE}
\end{equation}
Here, $\psi' (x',t')\equiv \psi'_{1}=\psi'_{2}$ represents the wave function of both
components whereas
$\delta g=g_{12}+g$ describes the balance of mean-field intracomponent repulsion
and intercomponent 
attraction~\cite{petrov2015quantum,mistakidis2022cold}, 
%see Ref.~\cite{PhysRevA.101.051601} for further details on the rescaling of Eq.~(\ref{dreq1}). 
The effective
interaction strengths are experimentally tunable by dint of the
Fano-Feshbach~\cite{chin2010feshbach} or confinement-induced 
resonances~\cite{olshanii1998atomic}, 
see also Ref.~\cite{semeghini2018self} for the
magnetic-field dependence of the scattering lengths in $^{39}$K.
The energy of the system is
measured in units of $\hbar ^{2}/(m\xi ^{2})$, where $\xi =\pi \hbar ^{2}\sqrt{|\delta g|}/(mg\sqrt{2g})$ is the healing length. Rescaling time, length, and
wave function as $t'=\hbar/\left(m\xi ^{2}\right) $, $x'=\xi x$ and 
$\psi'=(2\sqrt{g})^{3/2} \psi/(\pi \xi (2|\delta g|)^{3/4})$, respectively, Eq.~(\ref{dGPE}) can be cast in the 
dimensionless form
\begin{equation}
i\psi_{t}=-\frac{1}{2}\psi_{xx}+|\psi |^{2}\psi -|\psi |\psi .
\label{dreq1}
\end{equation}

In the following, we focus on the stability and interactions of both  droplets
and bubbles. It is the balance between the residual mean-field repulsion,
LHY attraction and the kinetic energy which allows the self-trapping of
stable droplets in free space, see also Ref.~\cite{PhysRevA.101.051601}.
While our findings, as reported below, persist in the presence of a weak
external trap, here we restrict the discussion to the most fundamental
setting that does not include an external potential. Typical evolution times
of $t\sim 10^{3}$ in the scaled units, that we consider here, translate, in a
typical experimental setup~\cite{katsimiga2020observation}, with the
perpendicular trap $\omega_{\perp }=2\pi \times 250$~Hz, into $600$ ms in
physical units.

%%%%%%%%%%%%%%%%%%%%%%%%%%%%%%%%%%%%%%%%%%%%%%%%%%%%%%%%%%%%%%%
\begin{figure}[tbp]
\includegraphics[height=7.6cm]{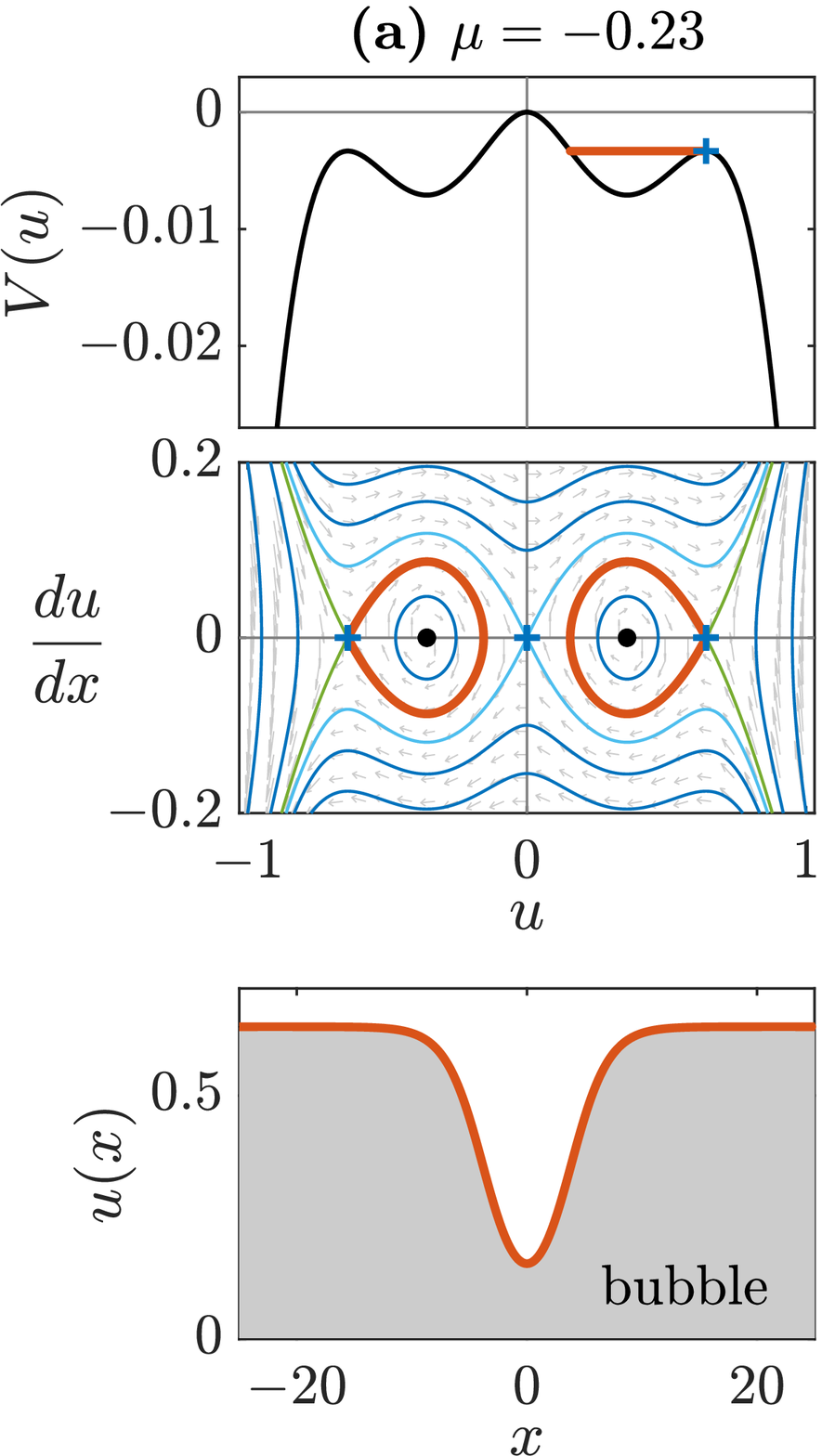} 
~~
\includegraphics[height=7.6cm]{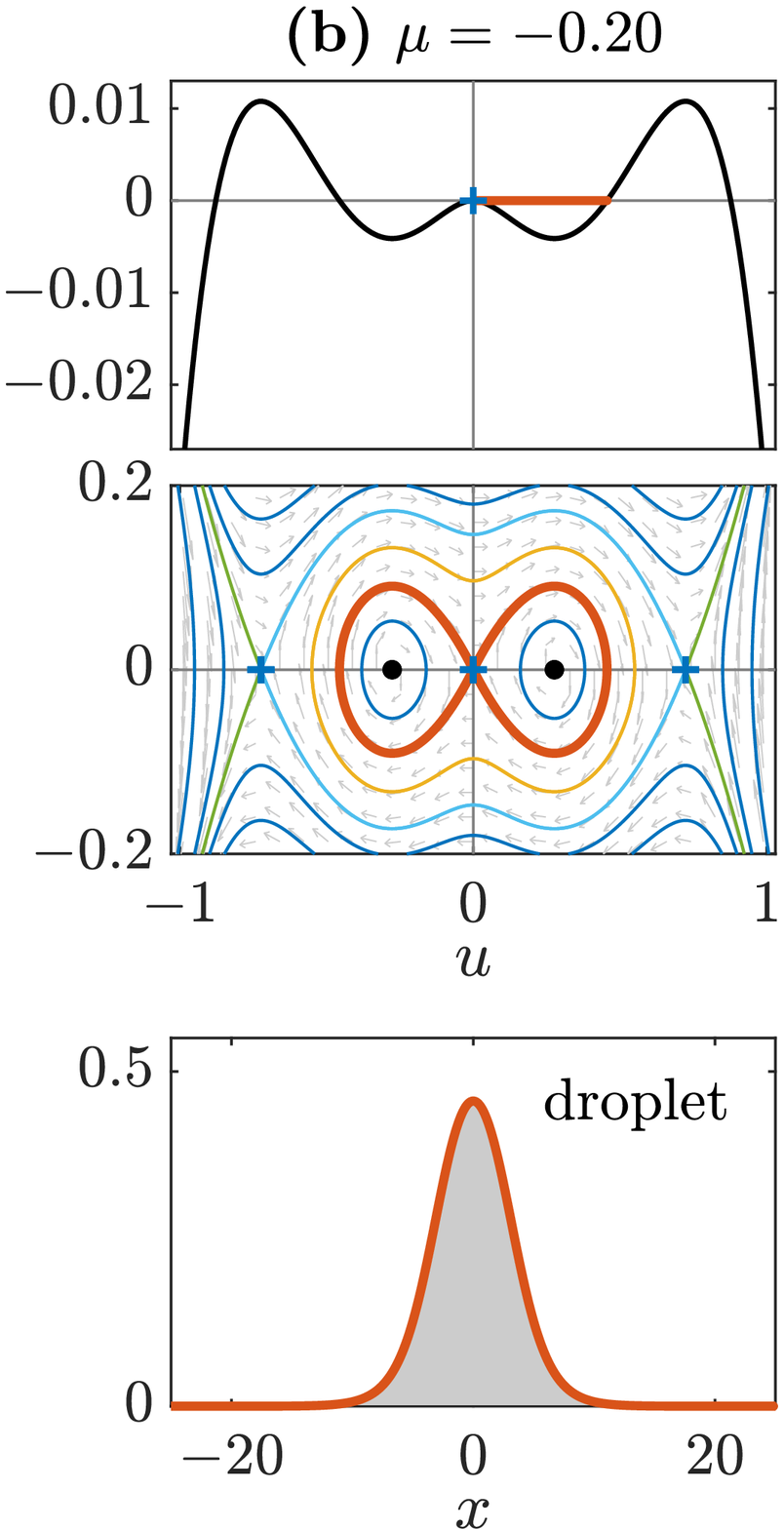}
\caption{
Self-localized nonlinear structures (see thick red curves) supported by the LHY correction.
Typical (a) bubble and (b) droplet solutions (bottom panels) corresponding to two types 
of homoclinic connections (middle panels) stemming from the Newtonian effective potential reduction (top panels). 
The bubble refers to a homoclinic connection anchored at the non-zero
$u_{+}$ constant background while the droplet represents a homoclinic 
connection of the zero ($u_{\ast}=0$) background.
}
\label{fig1}
\end{figure}
%%%%%%%%%%%%%%%%%%%%%%%%%%%%%%%%%%%%%%%%%%%%%%%%%%%%%%%%%%%%%%%

%%%%%%%%%%%%%%%%%%%%%%%%%%%%%%%%%%%%%%%%%%%%%%%%%%%%%%%%%%%%%%%
\subsection{The effective-potential picture and existence regions of
droplets and bubbles}
%%%%%%%%%%%%%%%%%%%%%%%%%%%%%%%%%%%%%%%%%%%%%%%%%%%%%%%%%%%%%%%

When seeking for stationary-state solutions of Eq.~(\ref{dreq1}), the usual
ansatz $\psi (x,t)=e^{-i\mu t}u(x)$ is employed, leading to the second-order ordinary differential equation (ODE) 
---or 2D (Hamiltonian) dynamical system---
\begin{equation}
\frac{d^{2}u}{dx^{2}}=2|u|^{2}u-2|u|u-2\mu u,  \label{dreq2}
\end{equation}
where $x$ plays the role of 
the evolution variable~\cite{katsimiga2023solitary}.
Without loss of generality for our 1D problem, we may assume that the stationary solutions are
real, reducing Eq.~(\ref{dreq2}) to the Newtonian equation of motion for a
unit-mass particle with coordinate $u$ experiencing the effective potential
\begin{equation}
V(u)=\mu u^{2}+\frac{2}{3}u^{2} |u|-\frac{1}{2}u^{4}.   
\label{epot}
\end{equation}
Characteristic profiles of $V(u)$ are shown in Fig.~\ref{fig1} for two 
chemical potentials. 
The fixed points of this potential correspond to
\begin{equation}
\label{dreq3}
u_{\ast }=0\text{~~and~~}u_{\pm }=\frac{1\pm \sqrt{1+4\mu }}{2}.  
\end{equation}
The homogeneous state associated with $u_{-}$ is modulationally 
unstable~\cite{mithun2019inter}, therefore we do not consider it here. 
On the other hand, both 
$u_{\ast }$ and the homogeneous state $u_{+}$ are modulationally stable,
with the latter arising at $\mu =-1/4$ through a saddle-center bifurcation (together with $u_{-}$), as seen in Eq.
(\ref{dreq3}). Then, the shape of the effective potential $V(u)$ defined by
Eq.~(\ref{epot}) determines the nature of the solitonic states that can
exist in this setting. Specifically, local maxima of $V(u)$ located at $\pm
u_{+}$ and $u_{\ast }$ have equal heights at $\mu =-2/9$.

Turning to the case of $-1/4<\mu <-2/9$, when the maximum of $V(u)$ at 
$u=u_{\ast }$ is higher than that at $u=\pm u_{+}$ [top panel of Fig.~\ref{fig1}(a)], Eq.~(\ref{dreq2}) produces a bubble (or its negative) homoclinic
[i.e., tending asymptotically to the same steady state
from the Greek ``$\kappa \lambda \acute{\iota} \nu \omega$'' (to tend) and 
``$\acute{o} \mu o \iota o$'' (same)]
to $u_{+}$, see red curves in the middle and bottom panels of Fig.~\ref{fig1}(a).
For chemical potentials in the interval 
$-2/9<\mu <0$, where the maxima at $\pm u_{+}$ are higher than the one at 
$u_{\ast }$ [top panel in 
Fig.~\ref{fig1}(b)], there exists a bright droplet [in addition to a heteroclinic
dark-soliton state that has been examined elsewhere~\cite{edmonds};
heteroclinic stems from ``$\kappa \lambda \acute{\iota} \nu \omega$'' 
and ``$\acute{\epsilon} \tau \epsilon \rho o$'' (different), i.e., tending 
asymptotically to different steady states], 
see red curves in the middle and bottom panels in Fig.~\ref{fig1}(b). 
For $\mu =0$, the homogeneous states $u_{-}$ and 
$u_{\ast }$ collide through a subcritical pitchfork 
bifurcation~\cite{katsimiga2023solitary} and, finally, at $\mu >0$ the 
equilibrium point at $u_{\ast }=0$ becomes a local minimum of the potential. In the latter case,
the situation is similar to that known for the cubic nonlinear 
Schr\"{o}dinger (NLS) equation and it is not considered further herein. As elaborated below, the
states of interest, namely bubbles and ``bright"
droplets exist, respectively, for $-1/4<\mu <-2/9$ and $-2/9<\mu <0$.
These chemical potential intervals will be our main focus,
in addition to the separatrix value of $\mu=-2/9$ where kinks
and antikinks exist.

%%%%%%%%%%%%%%%%%%%%%%%%%%%%%%%%%%%%%%%%%%%%%%%%%%%%%%%%%%%%%%%
\begin{figure*}[tbp]
\includegraphics[width=0.95\textwidth]{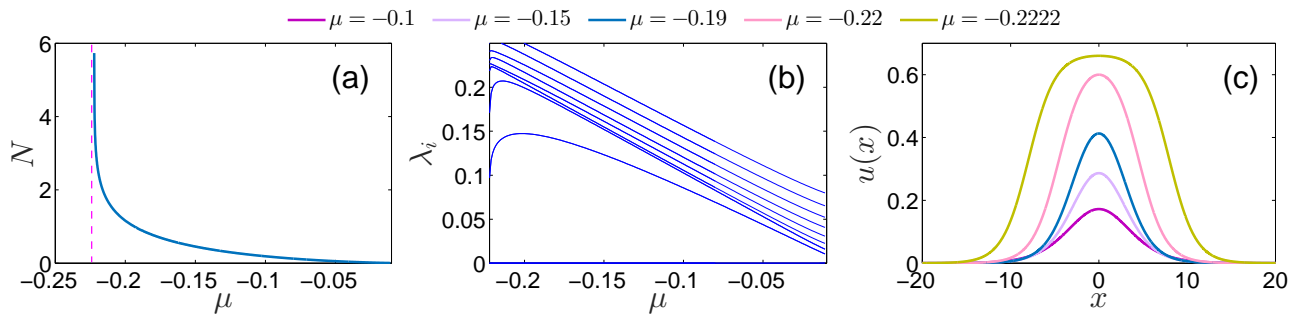}
\caption{(a) Numerically obtained scaled atom number for the droplet solution as a function of the chemical
potential. An increase of $N$ occurs when tending to the transition point below which only bubbles exist. This behavior is in accordance to the analytical prediction of Eq.~(\ref{dreq5}). 
(b) The imaginary part of the linearization eigenvalues, $\lambda_i$, is depicted with respect to $\mu$. 
Spectral stability, whereby $\lambda_{r}\equiv 0$, can be inferred independently of $\mu$ lying within the bright droplet regime.  
(c) Bright droplet waveforms for different values of $\mu$ (see legend) deforming from a quasi-Gaussian profile as $\mu \rightarrow 0^{-}$ to a flat-top configuration as $\mu \rightarrow -2/9$.   
These shapes and overall trend are in excellent agreement, as they should, with the exact solution of Eq.~(\ref{dreq4}).}
\label{fig2}
\end{figure*}
%%%%%%%%%%%%%%%%%%%%%%%%%%%%%%%%%%%%%%%%%%%%%%%%%%%%%%%%%%%%%%%

%%%%%%%%%%%%%%%%%%%%%%%%%%%%%%%%%%%%%%%%%%%%%%%%%%%%%%%%%%%%%%%
\subsection{The Bogoliubov-de Gennes linearized equations}
%%%%%%%%%%%%%%%%%%%%%%%%%%%%%%%%%%%%%%%%%%%%%%%%%%%%%%%%%%%%%%%

To address the stability of  droplets and bubbles, Eq.~(\ref{dreq1}) is
linearized for small perturbations ($f,g$) around the respective stationary
solution $u_{0}(x)$.
This is done via the ansatz:
$$
\psi(x,t)=u_0(x)+ \epsilon \left[f(x,t) + i g(x,t) \right],
$$
yielding the following Bogoliubov-de Gennes (BdG) equations  
\begin{subequations}
\label{dreq67}
\begin{eqnarray}
\frac{\partial f}{\partial t} &\!\!=\!\!&L_{1}g\equiv \left[ -\frac{1}{2}
\frac{\partial ^{2}}{\partial x^{2}}-F(u_{0}^{2})\right] g,  \label{dreq6} \\
\frac{\partial g}{\partial t} &\!\!=\!\!&-L_{2}f
%\equiv -\left[ -\frac{1}{2} \frac{\partial ^{2}}{\partial x^{2}}
%-F(u_{0}^{2})-2u_{0}^{2}F^{\prime }(u_{0}^{2})
\equiv \left[ \frac{1}{2} \frac{\partial ^{2}}{\partial x^{2}}
+F(u_{0}^{2})+2u_{0}^{2}F^{\prime }(u_{0}^{2})
\right] f ,\qquad  \label{dreq7}
\end{eqnarray}
\end{subequations}
where $F(y)=-y+\sqrt{y}$. 
$L_{1}$ and $L_{2}$ are the self-adjoint linearization operators with domain 
$D(L_{1})=D(L_{2})=H^{2}(\mathbb{R})$ while the adopted notation was previously introduced,
e.g., in Refs.~\cite{baras1989,debouard}.  Using a separation of variables
of the eigenfunctions $(f,g)^T$ (where $^T$ denotes transposition),
and a temporal part proportional to $e^{\lambda t}$, leads to an eigenvalue
problem whose solution enables the consideration of the spectral
stability of the obtained solutions, as we will discuss in further detail below.

%%%%%%%%%%%%%%%%%%%%%%%%%%%%%%%%%%%%%%%%%%%%%%%%%%%%%%%%%%%%%%%
\section{Stability analysis for droplets}
\label{stability_brightdrop}
%%%%%%%%%%%%%%%%%%%%%%%%%%%%%%%%%%%%%%%%%%%%%%%%%%%%%%%%%%%%%%%

The droplet solution of Eq.~(\ref{dreq2}),
\begin{equation}
u_{\mathrm{\,droplet}}(x)=\frac{\frac{2}{3}\frac{\mu }{\mu_{0}}}{1+\sqrt{1-
\frac{\mu }{\mu_{0}}}\cosh \left( \sqrt{-2\mu }(x-x_{0})\right) },
\label{dreq4}
\end{equation}
with $\mu_{0}\equiv -2/9$, exists in the interval of 
$\mu_{0}<\mu <0$~\cite{PhysRevA.101.051601}. In the limit of $\mu =\mu_{0}$, the droplet turns
into a kink solution, which connects the asymptotically constant values, 
$u=0 $ and $u=2/3$, fixed at 
$x\rightarrow \pm \infty$~\cite{PhysRevA.101.051601,katsimiga2023solitary}:
\begin{equation}
u_{\mathrm{\,kink}}(x)=\frac{1}{3}\left[1+\tanh\left(\frac{x}{3}\right)\right].  \label{kink}
\end{equation}
The scaled number of atoms in the droplet, 
$N=\int_{-\infty }^{+\infty }u^{2}(x)~dx$, is
\begin{equation}
N_{\mathrm{droplet}}=\frac{4}{9}\sqrt{\frac{-2}{\mu_{0}}}\left[ \ln \left(
\frac{1+\sqrt{\frac{\mu }{\mu_{0}}}}{\sqrt{1-\frac{\mu }{\mu_{0}}}}\right)
-\sqrt{\frac{\mu }{\mu_{0}}}\right].  \label{dreq5}
\end{equation}
The latter is provided in  Fig.~\ref{fig2}(a) as a function of the chemical potential. 
Evidently, it diverges as $\mu \rightarrow \mu_{0}$ and decays to zero as 
$\mu \rightarrow 0^{-}$.

The BdG spectrum assuming the above-mentioned separation of
variables, namely $f=e^{\lambda t} \tilde{f}$
and $g=e^{\lambda t} \tilde{g}$ in Eqs.~(\ref{dreq67}) with eigenvalues  
$\lambda=\lambda_r + i \lambda_i$ is presented in Fig.~\ref{fig2}(b). 
Notice that all eigenvalues remain imaginary confirming the spectral stability of the bright droplet throughout its region of existence. 
We do not dwell on the relevant
result and the bifurcation of internal modes of the bright droplet stemming from the 
band edge of the continuous spectrum at $\lambda_i=\mu$ since the relevant features have been 
analyzed in Ref.~\cite{PhysRevA.101.051601}. 
Characteristic profiles of the bright droplet solution for varying 
values of $\mu$ are shown in Fig.~\ref{fig2}(c). 
The bright droplet changes its shape from a Gaussian ($\mu \rightarrow 0^{-}$) to a flat-top ($\mu \rightarrow \mu_0^+$) configuration for decreasing $\mu$.

%%%%%%%%%%%%%%%%%%%%%%%%%%%%%%%%%%%%%%%%%%%%%%%%%%%%%%%%%%%%%%%
\begin{figure*}[tbp]
\includegraphics[width=1.0\textwidth]{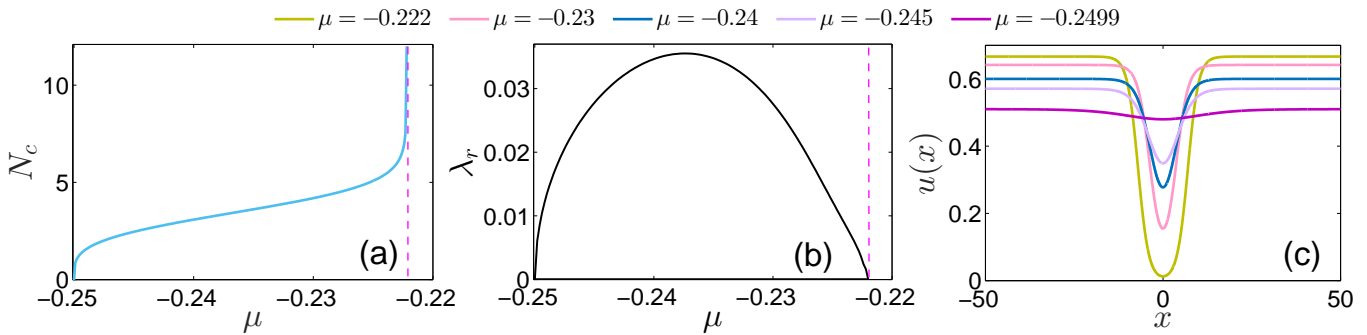}
\caption{(a) Complementary scaled atom number, $N_c$, see also
Eq.~(\ref{Nc}), for a bubble stationary state in terms of $\mu$. 
(b) The real part of the BdG spectrum upon chemical potential variation demonstrating the unstable nature of bubbles throughout their domain of existence. 
Dashed magenta lines in both (a) and (b) mark the boundary above which bubbles cease to exist.  
(c) The wave function of bubbles for distinct $\mu$ values (see legend). 
As it can be seen, for smaller $|\mu|$, 
the bubble's core becomes deeper, tending to be infinitely broad in the
limit of $\mu \rightarrow -1/4$. }
\label{fig3}
\end{figure*}
%%%%%%%%%%%%%%%%%%%%%%%%%%%%%%%%%%%%%%%%%%%%%%%%%%%%%%%%%%%%%%%

%%%%%%%%%%%%%%%%%%%%%%%%%%%%%%%%%%%%%%%%%%%%%%%%%%%%%%%%%%%%%%%
\section{Stationary bubbles}
\label{stability_bubble}
%%%%%%%%%%%%%%%%%%%%%%%%%%%%%%%%%%%%%%%%%%%%%%%%%%%%%%%%%%%%%%%

We subsequently investigate bubbles appearing  
for $-1/4< \mu < \mu_0=-2/9$, see also Fig.~\ref{fig1}(a).
Utilizing the first integral formulation of the relevant
quadrature problem of Eqs.~(\ref{dreq2})--(\ref{dreq3}), 
the relevant solution can be obtained via  
\begin{equation}
\int \frac{du}{\sqrt{2(E-V(u))}}=\pm (x-x_{0}).  \label{dreq12}
\end{equation}
The bubble is represented by the homoclinic orbit attached to $u=u_{+}$,
with mechanical energy $E=V(u_{+})$ [see Eq.~(\ref{epot})], which can be
obtained through the substitution $w=u-u_{+}$. A direct calculation, as
expected, shows that the quantity under the radical in Eq.~(\ref{dreq12})
can be rewritten in a quadratic form with respect to $w$,
$$
E-V(w)=\frac{w^{2}}{2}\left( A+Bw+w^{2}\right),
$$
where $A\equiv 1+4\mu +\sqrt{1+4\mu }>0$ and 
$B\equiv 2(1+3\sqrt{1+4\mu})/3>0$. The evaluation of the integral, see Ref.~\cite{Grad}, leads to the
exact analytical expression for a bubble centered at $x=x_0$
\begin{equation}
u_{\rm \,bubble}(x)=u_{+}-\frac{2A}{B+C\cosh \big(\sqrt{A}(x-x_{0})\big)}
,\quad   \label{dreq14}
\end{equation}
with $C\equiv \sqrt{B^{2}-4A}$.
The \textit{complementary mass} of the bubble,
\begin{equation}
N_{c}\equiv \int_{-\infty }^{+\infty }\big(u_{+}^{2}-u^{2}(x)\big)dx,
\label{Nc}
\end{equation}
is shown as a function of $\mu $ in Fig.~\ref{fig3}(a). 
As it can be seen, $N_c \to 0$ as $\mu \to -1/4$ since the
relevant bubble profile becomes nearly homogeneous [see Fig.~\ref{fig3}(c) for $\mu=-0.2499$ just
before the branch disappearance] while it diverges in the limit of 
$\mu\rightarrow \mu_0=-2/9$ where the bubble core is the deepest.
The progressive modification of this solution towards a homogeneous (flat) one as $\mu$ increases is also presented in Fig.~\ref{fig3}(c). 
The above-described trend of $N_c$ is reminiscent of the one found for bright droplets but in terms of $N$ [Fig.~\ref{fig2}(a)].

The underlying stability properties of a single bubble upon chemical potential variations are depicted in Fig.~\ref{fig3}(b). 
Specifically, only the real eigenvalues, $\lambda_r$, versus $\mu$ are illustrated demonstrating the spectral {\em instability} of this configuration throughout its interval of existence. 
For instance, the unstable eigenvalues
for the limit bubble solutions, i.e., for $\mu =-0.2499$ and $\mu =-0.2222$
are $(\lambda_{r},\lambda_{i})=(0.0276,0)$ and $(\lambda_{r},\lambda
_{i})=(0.0043,0)$, respectively. 
We remark that the time at which the instability dynamically manifests itself is inversly proportional to the magnitude of the
ensuing eigenvalue.
Our findings are supported by the results of Refs.~\cite{baras1989,debouard} arguing that
in NLS settings bearing multi-stability, bubble-like  states when they exist are always unstable.

Applying the arguments of Refs.~\cite{baras1989,debouard} to the present
case, we note the following. The stationary wave function $u_{0}$ satisfies 
$L_{1}u_{0}=0$, however $u_{0}\notin L^{2}(\mathbb{R})$, given the
asymptotics of the bubble. Accordingly, $0$ is not an eigenvalue of $L_{1}$,
hence the range of $L_{1}$ is dense in $L^{2}(\mathbb{R})$ and we can define
the inverse operator $L_{1}^{-1}$ as an unbounded one with a dense domain in
$L^{2}(\mathbb{R})$~\cite{baras1989,debouard}. Indeed Ref.~\cite{debouard}
proves in Lemma 3.2 that $L_{1}$ is a positive operator, i.e., $(L_{1}u,u)>0$
holds for all $u\in H^{1}(\mathbb{R})$. On the other hand, it is
straightforward to use the Sturm-Liouville theory to prove that $L_{2}$ has
a negative eigenvalue. This is because the derivative of the bubble, 
$du_{0}/dx$, is an odd function in $L^{2}(\mathbb{R})$ with a single node
satisfying $L_{2}\left( du_{0}/dx\right) =0$. Hence, the Sturm comparison
theorem for this self-adjoint operator establishes the existence of a
nodeless negative eigenfunction. The combination of the two statements leads
to Theorem 3.1 in Ref.~\cite{debouard}, which suggests that in that case the
eigenvalue problem represented by Eqs.~(\ref{dreq67}) can be written as
\begin{eqnarray}
\lambda ^{2}L_{1}^{-1}f &=&-L_{2}f   \notag \\
\Rightarrow~~\lambda ^{2} &=&-\inf_{u\in H^{1}(\mathbb{R})\cap D(L_{1}^{-1})}
\frac{\langle L_{2}f,f\rangle }{\langle L_{1}^{-1}f,f\rangle }.  \label{dreq15}
\end{eqnarray}
Furthermore, one can establish an upper bound of the relevant eigenvalue as 
$-\lambda ^{2}<\beta ^{2}/4$, where $\beta =\sup [-2u_{0}^{2}(x)F^{\prime
}(u_{0}^{2}(x)]$~\cite{debouard}. The corresponding existence of an
eigenvector such that the fraction in Eq.~(\ref{dreq15}) is negative
confirms the existence of an associated eigenvalue pair with $\lambda ^{2}>0$
which, in turn, renders the bubble generically unstable. It is interesting
to contra-distinct the present case with the solitary waves vanishing at 
$x\rightarrow \pm\infty $, in which case $0$ \textit{is} an eigenvalue of 
$L_{1}$~\cite{Kapitula2013}. This point explains the difference
between the instability of the bubbles and the stability of the droplets.
An interpretation of such instability
provided, e.g., in the work of Ref.~\cite{baras1989} 
has to do with the metastability of the relevant
configuration, in comparison with the global
minimum of the energy, involving 
the homogeneous state with vanishing
amplitude. This is reflected also in the
unstable dynamics of the state to which we
will return in the next section. However,
the difference between that and the 
stable dark soliton configuration lies
in the effective topological protection of
the latter (due to its phase jump) that
precludes its destabilization, contrary
to what we will see below for the bubbles.

%%%%%%%%%%%%%%%%%%%%%%%%%%%%%%%%%%%%%%%%%%%%%%%%%%%%%%%%%%%%%%%
\begin{figure*}[tbp]
\includegraphics[width=1.0\textwidth]{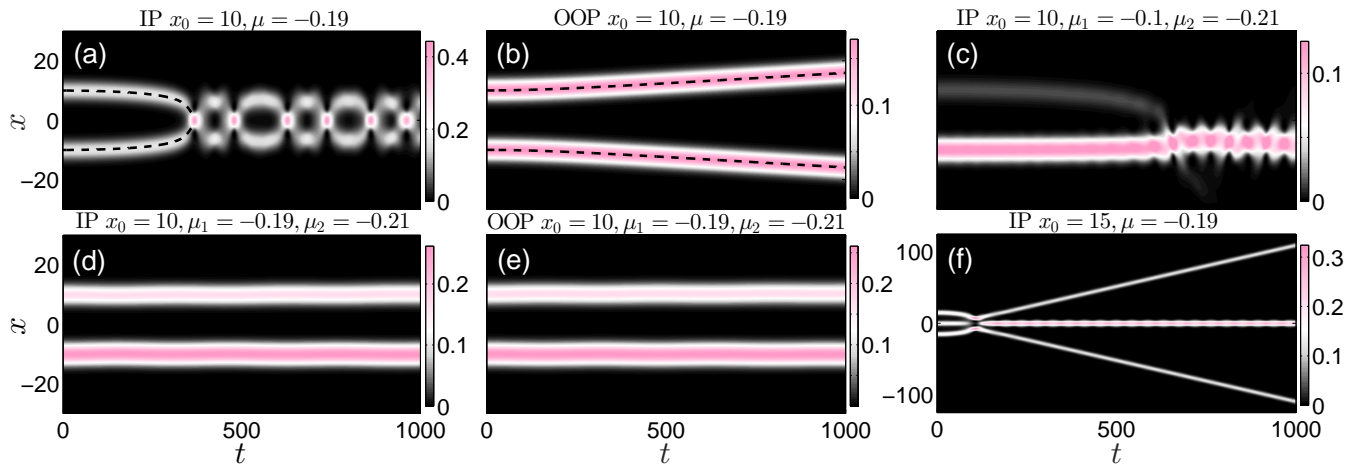}
\caption{(a)--(e) Spatiotemporal evolution of the density of two bright droplets, initially at rest, placed symmetrically with respect to $x=0$ at $x=\pm x_0$, with different relative phases ($\Delta \theta =0$ or $\pi$), and chemical potentials $\mu_1$, $\mu_2$ (see legends). 
It is evident that (a) in-phase (IP) bright droplets bearing the same $\mu$ feature attraction, whilst (b) out-of-phase (OOP) ones repel. 
In contrast, particle-imbalanced droplets either being (d) IP or (e) OOP remain nearly intact for small chemical potential differences. 
(c) Highly imbalanced IP droplets experience enhanced particle transfer.  
Dashed black lines indicate the prediction of the respective Manton's method which is in excellent agreement with the 
%amended GPE simulations. 
simulations of Eq.~(\ref{dreq1}).
(f) Density dynamics of three identical IP bright droplets leading to a central, heavier and excited droplet and two lighter
counter-propagating ones.}
\label{fig4}
\end{figure*}
%%%%%%%%%%%%%%%%%%%%%%%%%%%%%%%%%%%%%%%%%%%%%%%%%%%%%%%%%%%%%%%

%%%%%%%%%%%%%%%%%%%%%%%%%%%%%%%%%%%%%%%%%%%%%%%%%%%%%%%%%%%%%%%
\section{Dynamics and interactions of droplets}
\label{dynamics}
%%%%%%%%%%%%%%%%%%%%%%%%%%%%%%%%%%%%%%%%%%%%%%%%%%%%%%%%%%%%%%%

%%%%%%%%%%%%%%%%%%%%%%%%%%%%%%%%%%%%%%%%%%%%%%%%%%%%%%%%%%%%%%%
\subsection{Droplet collisions and Manton's method}
%%%%%%%%%%%%%%%%%%%%%%%%%%%%%%%%%%%%%%%%%%%%%%%%%%%%%%%%%%%%%%%

Next, we explore the dynamics of multiple droplets with phase differences 
$\Delta \theta $ ranging from $\Delta \theta =0$ (in-phase) to $\Delta \theta
=\pi $ (out-of-phase) in analogy with what has been experimentally probed
for bright solitons~\cite{borisrandy}. To examine the interactions between
identical in- and out-of-phase droplets, we develop the Manton's method~\cite{manton1979effective}
following the analysis reported in Ref.~\cite{mantonpgk} for the NLS model.
This method deploys the conservation laws for the scaled atom number, $N$, as well 
as the linear momentum,
\begin{equation}
P=\frac{i}{2}\int_{-\infty }^{+\infty }\big(uu_{x}^{\star }-u^{\star }u_{x}
\big)dx.  \label{dreq8}
\end{equation}
The basic idea is that, given the conservation laws associated with $N$ and $P$,
the calculation of their rates of change will result in contributions
stemming purely from the surface terms produced by the 
%integral expressions, such as the one in Eq.~(\ref{dreq8}). 
corresponding fluxes within the respective conservation laws.
Therefore, evaluating $dP/dt$ in a spatial domain between points $a$ and $b$, 
and using the equation of motion, leads to
\begin{equation}
\frac{dP}{dt}=\frac{1}{4}\bigg[uu_{xx}^{\star }+u^{\star }u_{xx}-2|u_{x}|^{2}
\bigg]_{a}^{b}.  \label{dreq9}
\end{equation}
On the other hand, $dP/dt$ amounts to a force. Then, if one assumes the
presence of two droplets centered at $x_{0}=0$ and $x_{0}=s$ and considering  
$a<0<b\ll s$, with sufficiently large $|a|$ and $b$, the expression in
Eq.~(\ref{dreq9}) mostly garners a contribution from the right boundary, since 
the left one, at $x=a$, is exponentially weaker. As such, the only
``force" at $x=b$ stems from the
interaction of the droplet at $x_{0}=0$ with that at $x_{0}=s$. Utilizing
the asymptotic form of the two droplets, namely 
$u_{1}=\eta e^{-\sqrt{-2\mu }x}$ and $u_{2}=\eta e^{i\Delta \theta }e^{\sqrt{-2\mu }(x-s)}$, for the
region near $x=b$, and using the ansatz $u=u_{1}+u_{2}$ 
[where $\Delta \theta $
is the phase shift between the two droplets and 
$\eta \equiv \mu /(3\mu_{0} \sqrt{1-\mu /\mu_{0}})$], 
one arrives at the effective force
\begin{equation}
\frac{dP}{dt}=-4\mu \eta ^{2}\cos (\Delta \theta )e^{-\sqrt{-2\mu }s}.
\label{dreq10}
\end{equation}
Subsequently, following the arguments of Ref.~\cite{mantonpgk}, it is
possible to derive the following evolution equation for the relative position between
the droplets:
\begin{equation}
\frac{d^{2}s}{dt^{2}}=-\frac{2}{N}\frac{dP}{dt}.  \label{dreq11}
\end{equation}
Here, the separation, $s$, between the two droplets has been
taken into account, as well as the fact that the mass of each droplet is
given by Eq.~(\ref{dreq5}), while the force takes the form of Eq.~(\ref{dreq10}).
Notice that the relative phase appears to play a similar role
as in the case of regular bright solitary waves~\cite{borisrandy,mantonpgk}, \textit{viz}., 
in-phase interaction leads to attraction, while an out-of-phase
one leads to repulsion. A similar approximation for the interaction between 
2D and 3D solitons was elaborated in 
Ref.~\cite{InteractionPotential}.
%%%%%%%%%%%%%%%%%%%%%%%%%%%%%%%%%%%%%%%%%%%%%%%%%%%%%%%%%%%%%%%
\begin{figure*}[tbp]
\includegraphics[height=7.6cm]{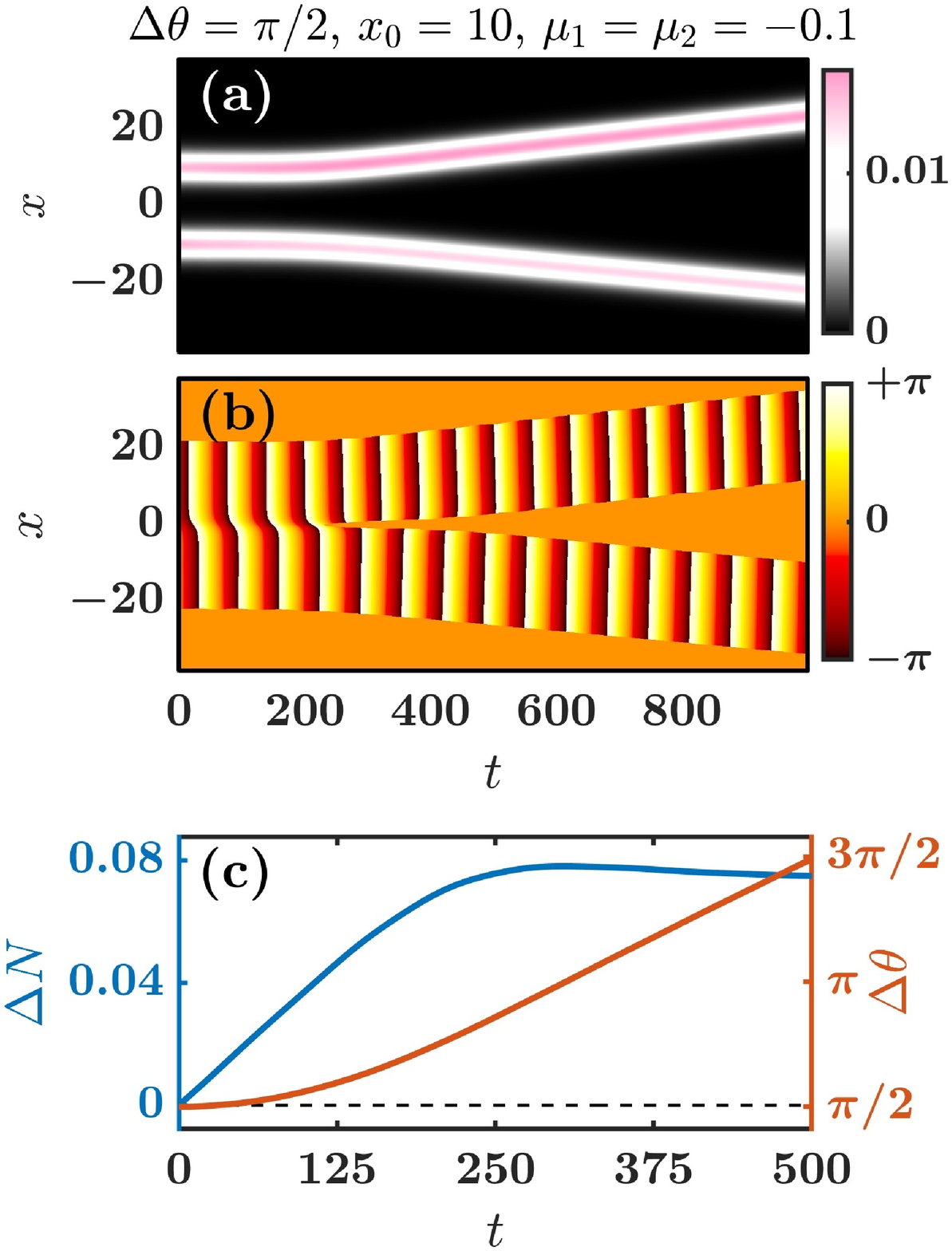}
\includegraphics[height=7.6cm]{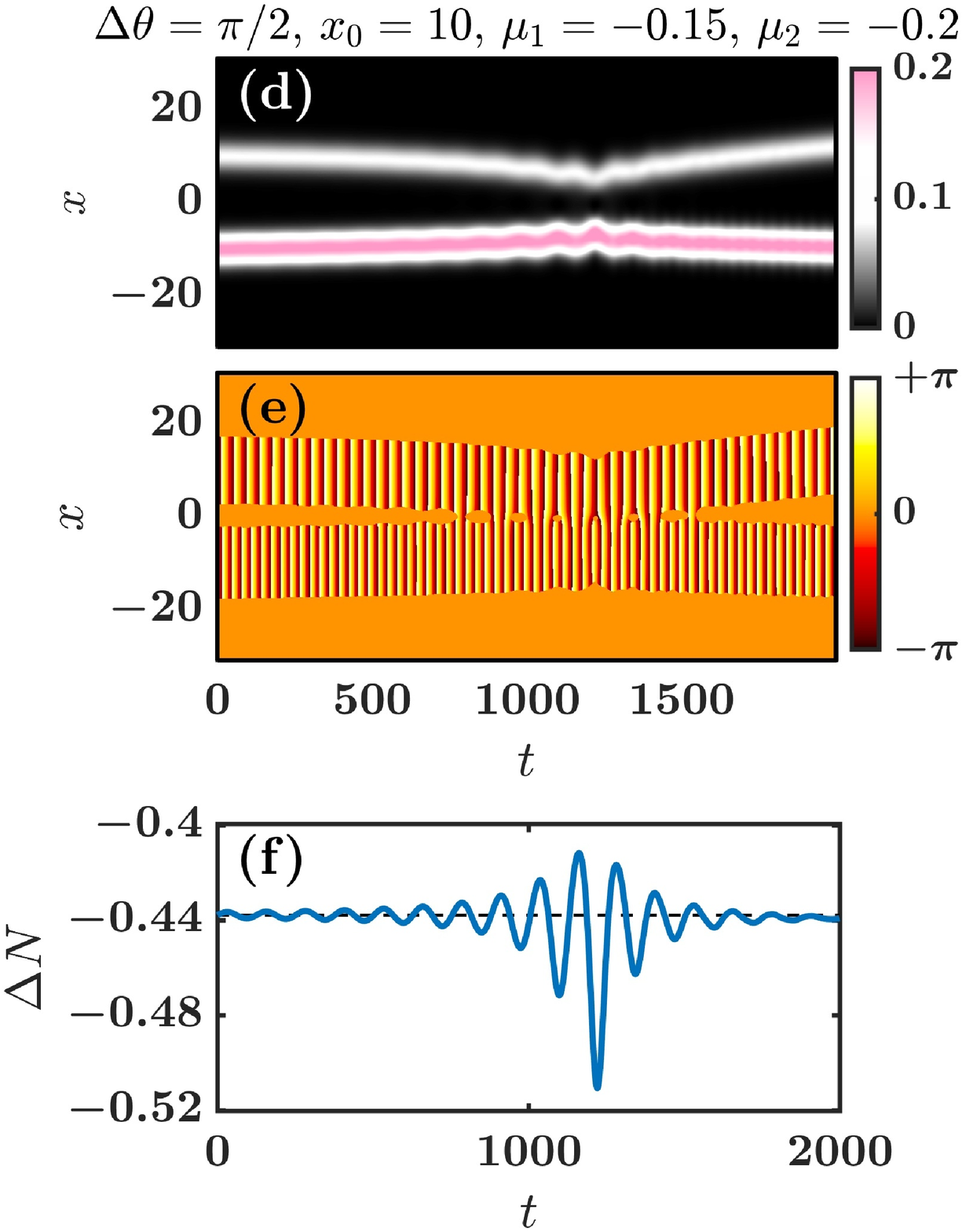}
\includegraphics[height=7.6cm]{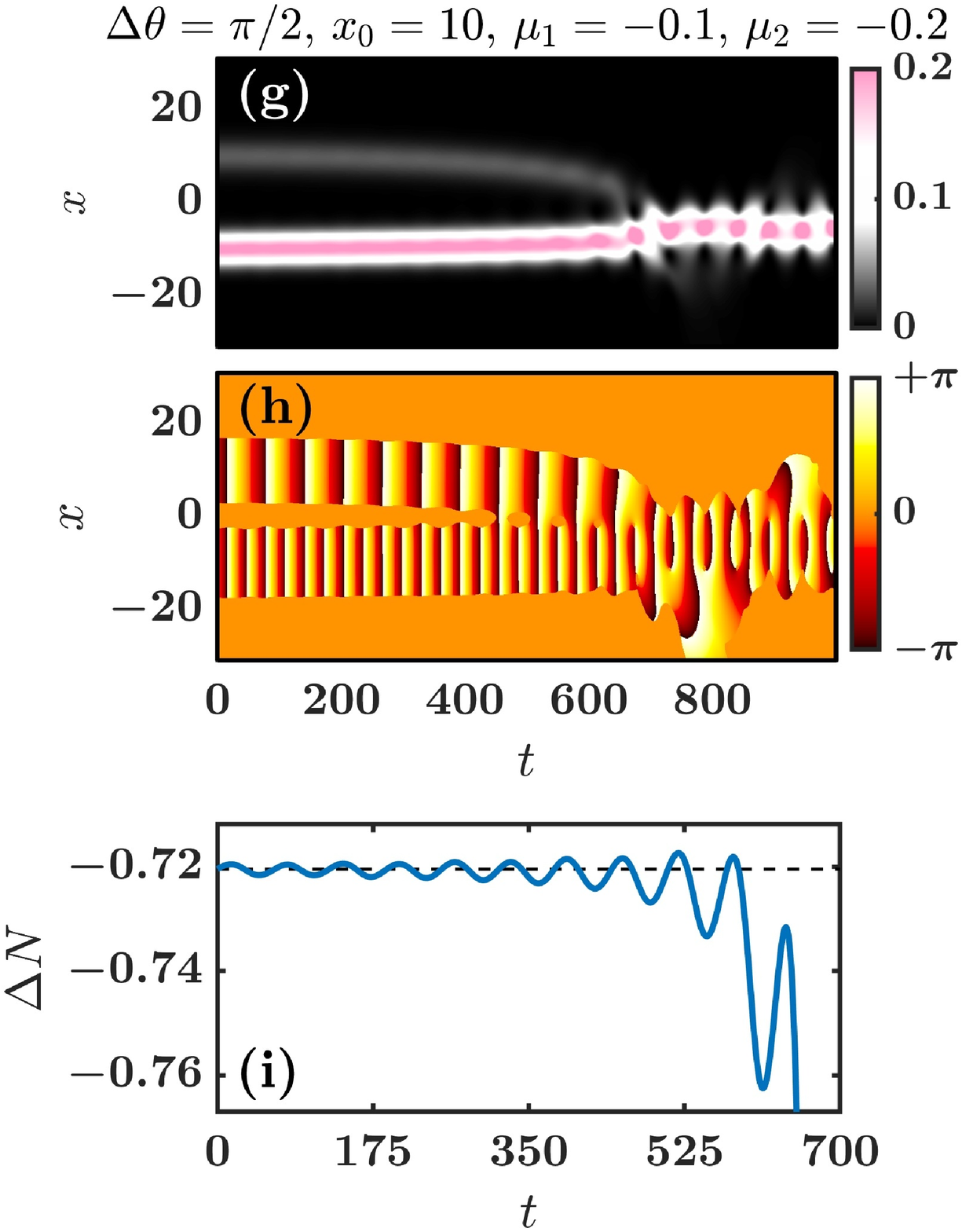}
\caption{Temporal evolution of two interacting droplets possessing a phase difference 
$\Delta \theta=\pi/2$ and initial separation $\Delta x_0 \equiv 2x_0=20$. 
Panels (a) and (b) depict, respectively, the spatio-temporal evolution 
of the density, $|\psi (x,t)|^2$, and phase of two identical droplets with $\mu_1=\mu_2=-0.1$ while panel (c) presents the corresponding time-evolution 
of the mass imbalance $\Delta N(t)$ [see Eq.~(\ref{eq:DN})] and
the relative phase, $\Delta\theta(t)$, between the two droplets.
(d)--(f) The same as (a)--(c) but for the case where the
initially seeded droplets have different chemical potentials, i.e. $\mu_1=-0.15$ and $\mu_2=-0.2$. 
(g)--(i) Similar to (d)--(f)  but for larger chemical potential differences, 
namely $\mu_1=-0.1$ and $\mu_2=-0.2$. 
In all cases, transfer of mass among the droplets takes place. 
}
\label{fig5}
\end{figure*}
%%%%%%%%%%%%%%%%%%%%%%%%%%%%%%%%%%%%%%%%%%%%%%%%%%%%%%%%%%%%%%%

Commenting more generally on the
applicability of the method, we note
the following.
In principle, Manton's method can be applied
when the model possesses an invariance with
respect to translation that leads to the
existence of a momentum conservation law.
In addition, the model needs to possess
solitary waves that feature an exponential
tail and effective particle dynamics, as is the case with numerous 
one-dimensional models that bear solitons
or solitary waves or even (exponentially
decaying in space) 
breather solutions. At the moment, we are
not familiar with applications of the (standard
Manton) method
in models that are not translationally 
invariant or ones in which the waves
feature power law tails. For the latter,
recently, a far more
elaborate method was introduced for capturing their interactions
in Ref.~\cite{Manton_2019}, yet to the best of our
knowledge, this has not been systematized
including in higher dimensional settings as of yet. The latter clearly constitutes
an interesting direction for further
explorations.

The dynamical prediction of Eq.~(\ref{dreq11}) has been tested for the case of in-phase (IP) bright droplet
attraction [Fig.~\ref{fig4}(a)], as well as in that of out-of-phase (OOP) droplet repulsion [Fig.~\ref{fig4}(b)]. 
Here, droplets are originally ($t=0$) placed at a relative distance $\Delta x_0=20$. 
The overall phenomenology is essentially the same for different 
$\Delta x_{0}$. 
It is evident that the time-evolved density, $|\psi (x,t)|^2$,  %$|u(x,t)|^2$, 
as captured by the amended GPE model of Eq.~(\ref{dreq1}) 
is in excellent agreement with the analytical estimate of the particle picture as concerns the droplet motion. This analytical estimate is obtained upon integrating Eq.~(\ref{dreq11}) and is directly compared with the attractive and the repulsive 
case, see dashed black lines in Fig.~\ref{fig4}(a) and (b) respectively. 
It is also relevant to point out that in the case where the
droplets attract, Eq.~(\ref{dreq11}) is utilized up to the point where the droplets lose their individual character by approaching
each other at distances comparable to their individual widths. 
In that regime, our assumption put forth for the Manton method
(in terms of $a \ll 0 \ll b \ll s$) is no longer valid and we
cannot aspire to describe the relevant dynamics accurately. 
Nevertheless, it is observed that in-phase droplets upon collision
attempt to separate, yet cannot escape each other's attraction
and collide anew for sufficiently long times, a pattern that recurs throughout the evolution. On the other hand, the repulsion of
OOP droplets leads them to indefinite separation.

%%%%%%%%%%%%%%%%%%%%%%%%%%%%%%%%%%%%%%%%%%%%%%%%%%%%%%%%%%%%%%%
\begin{figure*}[tbp]
\includegraphics[width=1.0\textwidth]{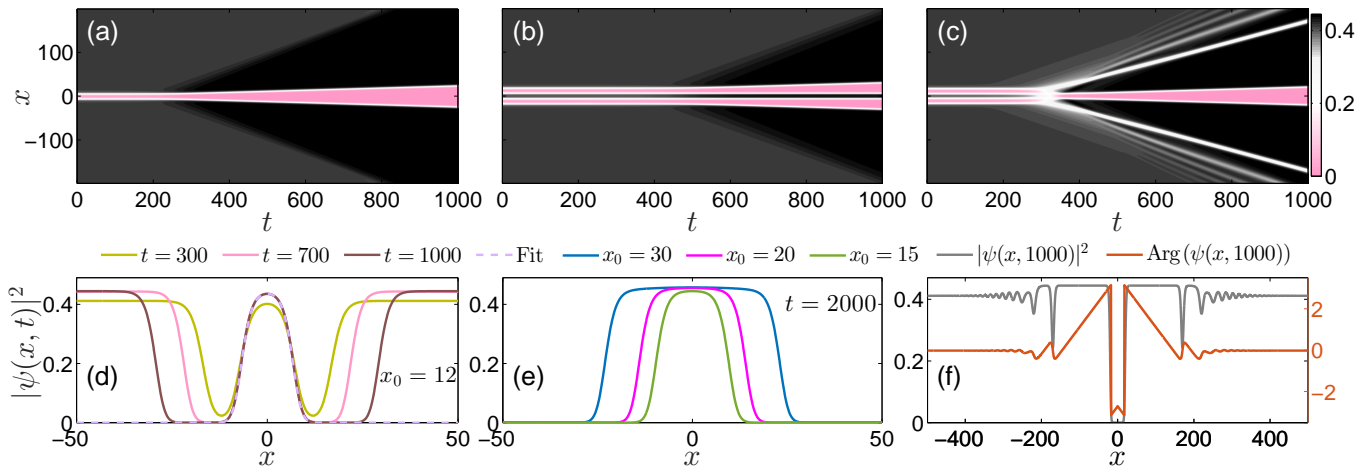}
\caption{(a) Density evolution of a single bubble for $\mu=-0.23$. 
The unstable dynamics of the bubble manifests through the expansion of its core. 
Dynamics of the density of two in-phase bubbles placed at a relative distance (b) $\Delta x_0=24$ and (c) $\Delta x_0=22$ exposing the crucial role of $\Delta x_0$. Namely, for relatively small $\Delta x_0$ the bubbles attract before destabilizing generating a heavier central bubble while emitting counter-propagating shock waves leaving behind gray solitons, see in panel (f) the corresponding profile and phase at $t=1000$.
Slightly larger distances lead to non-interacting bubbles whose cores expand in the course of the evolution trapping in between a droplet as illustrated in panel (d) at distinct time-instants for $\Delta x_0=24$ and in (e) at a specific time-instant and varying initial separation (see legends). Dashed line in panel (d) refers to a fit of the numerically obtained waveform to the analytical droplet solution given by Eq.~(\ref{dreq4}).}
\label{fig6}
\end{figure*}
%%%%%%%%%%%%%%%%%%%%%%%%%%%%%%%%%%%%%%%%%%%%%%%%%%%%%%%%%%%%%%%

To gain an overview of the droplet's dynamical response and interactions, situations 
where the two bright droplets
pertain to unequal chemical potentials, i.e., $\mu_1 \neq \mu_2$ are also explored. 
This scenario is shown in Figs.~\ref{fig4}(c), (d), and (e). 
Focusing on relatively small chemical potential differences depicted 
in Figs.~\ref{fig4}(d) and (e), e.g., $\mu_1=-0.19$ and $\mu_2=-0.21$, it is found that irrespectively of the relative phase ($\Delta \theta =0, \pi$) a quasi-stationary state occurs persisting for all evolution times ($t\sim 5 \times 10^3$). For instance, it can be seen that even for in-phase interactions with the two droplets initialized 
at the same distance as before, the relevant force is 
{\em much} weaker, not allowing to infer its definitive nature. 
Further increase of the chemical potential imbalance which is effectively 
related to the number-of-atoms, i.e., mass imbalance between the droplets, 
results in a drastically altered dynamics, see Fig.~\ref{fig4}(c). 
Particularly, the in-phase droplet collision entails a significant
transfer of mass from the lighter to the heavier one. 
Transfer of mass has been also reported in droplet collisions with finite relative velocity~\cite{Astrakharchik_2018}. 
Moreover, it turns out from the amended GPE predictions that imbalanced droplets experience weaker attraction than balanced ones as it can be inferred from the later occurrence of the relevant collision event, compare Figs.~\ref{fig4}(a) and (c). 
This inelastic process leads to the merger of the colliding objects into a single droplet in an excited state. 
Recall that droplet mergers can also be observed in a different context during the process of the so-called modulation instability mechanism~\cite{mithun2019inter,mithun2021statistical}.

As a next step, we investigate situations with a larger number of droplets.
A scenario of this sort is illustrated in Fig.~\ref{fig4}(f), consisting of a three droplet pair-wise in-phase initialization. 
It is found that the collision of the three identical (particle balanced) droplets appears to occur significantly faster compared to the two droplet case [Fig.~\ref{fig4}(a)] even though they are initially placed at larger distances. This is natural to expect
on the basis of the enhanced attraction arising in the present case.
Partial transfer of mass from the side droplets to the central one takes place producing two lighter and thus faster counter-propagating ones and a heavier central droplet in an excited state, which then exhibits breathing evolution. 
These findings motivate further studies and possible future 
work on the collisional dynamics of multi-droplet lattices, 
as such scenarios have been experimentally implemented
in the realm, e.g., of dark solitons in the work of Ref.~\cite{weller}.

%%%%%%%%%%%%%%%%%%%%%%%%%%%%%%%%%%%%%%%%%%%%%%%%%%%%%%%%%%%%%%%
\begin{figure*}[tbp]
\includegraphics[width=1.0\textwidth]{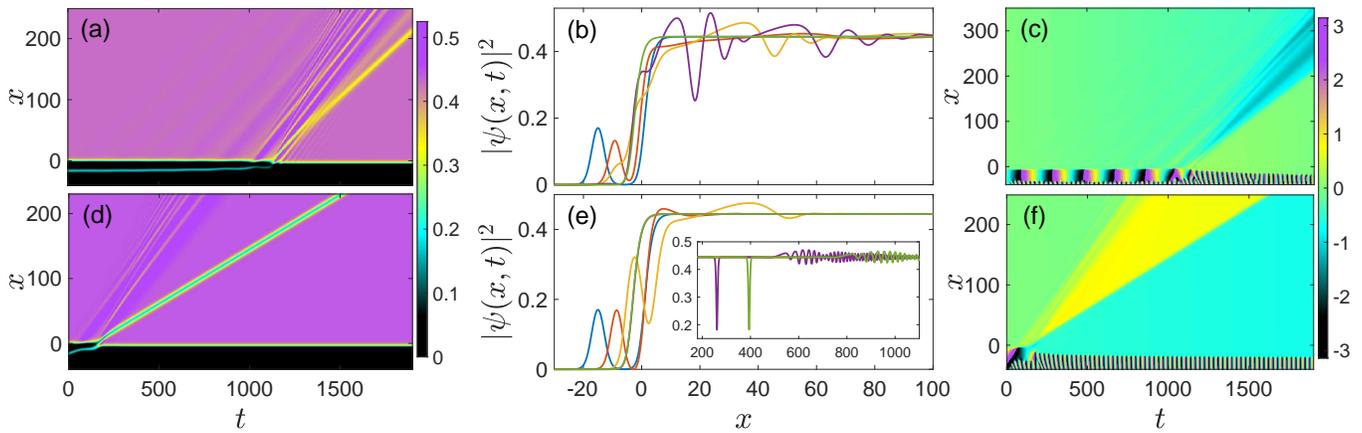}
\caption{(a) Density evolution for the interaction of a kink placed at $x=0$ with a bright 
droplet placed at $x_0=15$ while having (a) zero and (d) finite velocity $v=0.1$. 
(b) [(e)] Corresponding instantaneous profiles of (a) [(d)] at $t=0$ (blue line), $t=1023$ (orange line), $t=1128$ (yellow line), $t=1176$ (purple line) and $t=1800$ (green line) [$t=0$ (blue line), $t=68$ (orange line), $t=165$ (yellow line), $t=1200$ (purple line), and $t=1800$ (green line)]. 
(c) [(f)] Dynamics of the wave function's phase, ${\rm Arg}(\psi)$, for case (a) [(d)].  
The collision leads to the ``absorption" of the droplet by the kink accompanied by the excitation of the latter and the spontaneous emission of dispersive shock waves [see also the inset in panel (e)] leaving behind a robustly propagating gray soliton [captured by the characteristic phase jump in panel (f)].}
\label{fig7}
\end{figure*}
%%%%%%%%%%%%%%%%%%%%%%%%%%%%%%%%%%%%%%%%%%%%%%%%%%%%%%%%%%%%%%%

To further understand the role that the phase difference plays on droplet interactions, 
we additionally explored the dynamics for different values of the phase shift
between the two droplets [see Fig.~\ref{fig5}].
Phase differences between $0$ and $\pi/2$ manifest an initial attraction 
as predicted by the effective force of Eq.~(\ref{dreq10}) within the particle picture.
Interestingly, it is observed that the droplets experience a slow mass exchange
during their interaction. 
Note that the mass transfer among initially identical 
droplets implies symmetry breaking between them. An explanation for this effect in
soliton-soliton collisions, which can also be conceptually considered relevant to 
the present setting, was proposed in Ref.~\cite{Khaykovich}.
This involves the mismatch in the profiles' density centers and the mismatch in their phase centers,
leading, in turn, to the dynamical manifestation of a symmetry-breaking.
In our setting, this mass exchange is responsible for the modification of the chemical
potential of the individual droplets. Specifically, the droplet with the largest $|\mu|$
will advance its phase faster than 
the other one, see for instance Figs.~\ref{fig5}(e) and (h).
As a result, in the course of the evolution, $\Delta \theta (t)$ 
becomes larger than $\pi/2$ and eventually reaches $\Delta\theta=\pi$.  
Therefore, the droplets will experience repulsion.
A characteristic example where two droplets
with an initial phase difference $\Delta\theta=\pi/2$, corresponding
to a zero mutual force according to the particle picture, start exchanging mass
and, around $t\approx 250$, their phase difference gets closer to $\Delta\theta=\pi$
which leads to the droplets repelling each other is showcased in Fig.~\ref{fig5}(b), (c).
The inter-droplet mass exchange is quantified by defining the mass
imbalance parameter $-1\leq\Delta N\leq 1$ as follows:
%
%\begin{eqnarray}
%N_+(t)&=&\int_0^{+\infty} |u|^2\,dx,
%\quad
%N_-(t)=\int_{-\infty}^0 |u|^2\,dx,
%\nonumber
%\\[1.0ex]
%\Delta N(t)&=&\frac{N_+(t) - N_-(t)}{N_+(t) + N_-(t)}.
%\label{eq:DN}
%\end{eqnarray}
%
\begin{eqnarray}
N_+(t)&=&\int_0^{+\infty} |\psi|^2\,dx,
\quad
N_-(t)=\int_{-\infty}^0 |\psi|^2\,dx,
\nonumber
\\[1.0ex]
\Delta N(t)&=&\frac{N_+(t) - N_-(t)}{N_+(t) + N_-(t)}.
\label{eq:DN}
\end{eqnarray}  
Here, $\Delta N(t)=0$ corresponds to a mass-balanced scenario and
$\Delta N(t)>0$ ($\Delta N(t)<0$) to the right (left) droplet being more massive. 
Exploring other cases with $0<\Delta\theta<\pi/2$ (not shown for brevity),
an initial attraction was indeed observed for droplets initialized with
the same chemical potentials in accordance with the particle picture. 
However, the mass imbalance generally led to an eventual repulsion of the droplets. 
This phenomenology is absent for $0<\Delta\theta\ll 1$ 
where attraction is prevalent inducing droplet collisions
similarly to the recurrent collisional events observed for IP ($\Delta\theta=0$) 
identical (mass balanced) droplets presented
in Fig.~\ref{fig4}(a). 
Additionally, for $\pi/2<\Delta \theta<\pi$ repulsion dominates the dynamics (not shown here).

Paradigmatic droplet interactions where the droplets are initialized bearing different 
masses (i.e., distinct chemical potentials) are illustrated in Fig.~\ref{fig5}(d)--(f) 
for $\mu_1=-0.15, \mu_2=-0.2$ 
and Fig.~\ref{fig5}(g)--(i) 
with $\mu_1=-0.1, \mu_2=-0.2$.
In both cases 
oscillation of the mass imbalance [Fig.~\ref{fig5}(f), (i)] is accompanied by a weak
attraction between the droplets [Fig.~\ref{fig5}(d), (g)] until their collision where both processes are enhanced. 
Particularly, for increasing chemical potential imbalances a definitive collision point can be inferred leading to a perfect mass transfer and the generation of a merger, i.e. a heavier and excited single droplet as shown in Fig.~\ref{fig5}(g). 
However, in the case of smaller initial chemical potential imbalance, 
the droplets do not experience a clear collision point (i.e. they reach a minimum distance) before separating, see Fig.~\ref{fig5}(d). 
According to the above discussion the interaction dynamics
between droplets with $0<\Delta \theta<\pi/2$,  intrinsically involves mass exchange and 
strongly depends on the initial separation, relatives phases, and chemical potentials.
It is important to mention that the particle picture that was 
presented above is valid under the assumption of equal chemical
potentials between the two droplets. Therefore, if one is to cast a 
particle picture for non-mass-balanced droplets a different approach
is necessary and, indeed, worth developing.
Efforts to better understand more general droplet interactions, together
with their corresponding particle picture, fall outside of the scope
of the present work and will be reported elsewhere.

%%%%%%%%%%%%%%%%%%%%%%%%%%%%%%%%%%%%%%%%%%%%%%%%%%%%%%%%%%%%%%%
\subsection{Dynamics of bubbles and their interactions}
%%%%%%%%%%%%%%%%%%%%%%%%%%%%%%%%%%%%%%%%%%%%%%%%%%%%%%%%%%%%%%%

Having established the instability of the bubbles in their entire existence
interval [see Fig.~\ref{fig3}(b)], we proceed to explore their
instability-driven dynamics. To this end, we monitor the evolution of these
configurations in the framework of the amended GPE of  Eq.~(\ref{dreq1}), as presented in 
Fig.~\ref{fig6}(a) for the single bubble and Figs.~\ref{fig6}(b) and (c) for two
bubbles. Note that, when addressing configurations including two or more
bubbles, the necessity to match the backgrounds makes it possible to
consider solely bubbles with equal chemical potentials, so that they
asymptote to the same background value, $u_{+}$ [see Eq.~(\ref{dreq14})]. 
In this case, a natural ansatz is $\psi (x,t=0)=u_{0}(x-x_{0})\times
u_{0}(x+x_{0})/u_{+}$ where $x_{0}=\Delta x_{0}/2$ is half the separation
between the bubbles and $u_{0}(x)$ is given by the single-bubble analytical
solution as described by Eq.~(\ref{dreq14}).

The dynamics depicted in Fig.~\ref{fig6}(a) suggest that, in line with the
analysis of the general bubble dynamics provided in Refs.~\cite{baras1989,debouard},
the inherent instability experienced by a single bubble causes it 
to modify its width, eventually leading it to a continuous expansion.
%
%of the bubble. 
Naturally, to maintain the total-mass
conservation, the evolution leads to a slight increase of the height of the
background supporting the bubble.
It is interesting to note that this dynamical response is inherently related to the used Neumann or periodic boundary conditions, while it is naturally expected to be modified in the presence of hard-wall boundaries due to the fixed background.

On the other hand, in the case of two bubbles it is observed that the ensuing interaction depends crucially on their relative distance.  
Placing the bubbles sufficiently far apart, their individual instability arises before they have the opportunity to interact with each other [Fig.~\ref{fig6}(b)]. 
The individual core expansion persists for evolution times $t\sim 10^4$. 
However, it is not a symmetric expansion with respect to the center of each bubble's core. 
Rather, the outer part of each bubble extends as before,
while in the region in between the two bubbles, a density hump is developed and trapped. 
We were able, upon fitting, to identify the latter as a bright droplet, 
see in particular Fig.~\ref{fig6}(d). 
The chemical potential of the fitted droplet is $\mu=-0.2222$. Thus, the background density self-adjusts to the value corresponding to the droplet. 
The same dynamical bright droplet formation occurs upon further increase of the initial bubble separation as depicted in Fig.~\ref{fig6}(e). 
To shed further light on the bubble interactions we then bring them in sufficient proximity that enables their
interaction to commence early enough so that they merge, prior to 
the manifestation of their individual instability; see [Fig.~\ref{fig6}(c)]. 
Evidently, an apparent attraction occurs between
the two bubbles leading to a corresponding central merger and counter-propagating dispersive shock wave (DSW) structures with a downstream emission of a gray soliton train, see 
Fig.~\ref{fig6}(f)~\cite{EL201611}.  
This central bubble is in turn subjected to the previously discussed instability. 
It is also worthwhile to point out that the aforementioned emitted entities have nearly constant speed, as manifested by their
nearly constant slope in the space-time contour plots 
of Fig.~\ref{fig6}. 
This behavior can be inferred from their trajectories [see white
regions in Fig.~\ref{fig6}(c)]. 

%%%%%%%%%%%%%%%%%%%%%%%%%%%%%%%%%%%%%%%%%%%%%%%%%%%%%%%%%%%%%%%
\begin{figure*}[tbp]
\includegraphics[width=1.0\textwidth]{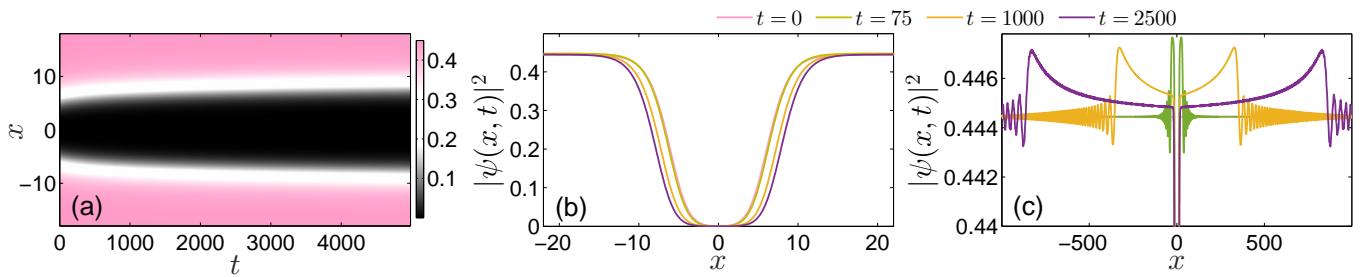}    
\caption{(a) Spatiotemporal evolution of the density, $|\psi(x,t)|^2$, in the course of
the apparently repulsive kink-antikink interaction with initial separation 
$\Delta x_{0}=10$. 
%Dashed lines in panel (a) mark the prediction of Manton's method for kink-antikink interactions.  
(b) Snapshots of (a) at different time-instants (see legend). (c) Magnification of (b) for visualizing the counter-propagating shock-wave emission.}
\label{fig8}
\end{figure*}
%%%%%%%%%%%%%%%%%%%%%%%%%%%%%%%%%%%%%%%%%%%%%%%%%%%%%%%%%%%%%%%

%%%%%%%%%%%%%%%%%%%%%%%%%%%%%%%%%%%%%%%%%%%%%%%%%%%%%%%%%%%%%%%
\subsection{Droplet-kink interactions}
%%%%%%%%%%%%%%%%%%%%%%%%%%%%%%%%%%%%%%%%%%%%%%%%%%%%%%%%%%%%%%%

Next, we address hybrid interactions between droplets and the kink, i.e.,
the exact stable state given by Eq.~(\ref{dreq4}) with that of Eq.~(\ref{kink}). The droplet is initially
placed at distance $x_{0}$ from the central point ($x=0$) of the kink.
Figure~\ref{fig7} displays examples of the ensuing interaction. In
particular, the top row of panels corresponds to an initial separation $x_{0}=15$ 
and zero initial velocity of the droplet. The latter initially features attraction that 
leads to its eventual collision and 
``absorption"\footnote{Partial reflection of the bright droplet that is of the order of $1\%$ occurs.} 
by the kink occurring around $t\sim 1100$; see the details below.
Figure~\ref{fig7}(b) offers a visualization of such collision and subsequent events imprinted in the instantaneous density profiles. The collision entails intrinsic excitation of the kink. In
particular, we observe spontaneous emission of a dispersive shock wave
traveling through the kink background and leaving behind a slower gray
soliton that remains intact, despite its interaction
with the former DSW while propagating. 
A characteristic signature of the gray soliton
is produced by the evolution of the phase, $\mathrm{Arg}(\psi )$, displayed
in Fig.~\ref{fig7}(c) which features a clear and constant phase jump. The
faint ripples in $\mathrm{Arg}(\psi )$, taking place ahead of the gray soliton
signature are shock-wave imprints. In general, the droplet gets ``absorbed" by
the kink, subsequently leading to the emission of the above
structures (DSW and gray soliton)
independently of the droplet's chemical potential (quantifying its
mass) and original position $x_{0}$. Naturally, the phenomenology happens
faster for heavier droplets and smaller distances (results not shown here).

This phenomenology can be generalized for droplets with initial velocity 
$v_{0}$. In this case, the collision is, as expected, accelerated, as seen
from the comparison of Figs.~\ref{fig7}(a) and (d), as well as of the corresponding 
snapshots in Figs.~\ref{fig7}(b) and (e). This conclusion is also supported
by monitoring the phase of the wave function, in the course of the
evolution, where signatures of both the shock wave (ripples) and the gray
soliton (phase jump being a fraction of $\pi$) can be distinguished. 
In this case, the emission of the dispersive shock wave 
and the nucleation of the gray soliton
become more conspicuous, see the inset of Fig.~\ref{fig7}(e) where both entities are
clearly captured. A more detailed study of collisions of traveling droplets
with kinks or other effective scattering potentials is a relevant direction
for studies beyond the scope of the present work. In this context, it will
be relevant to consider related transmission and reflection phenomena as it
was recently done in Ref.~\cite{debnath2023interaction} for a droplet colliding with a square well.

%%%%%%%%%%%%%%%%%%%%%%%%%%%%%%%%%%%%%%%%%%%%%%%%%%%%%%%%%%%%%%%
\subsection{The kink-antikink interaction}
%%%%%%%%%%%%%%%%%%%%%%%%%%%%%%%%%%%%%%%%%%%%%%%%%%%%%%%%%%%%%%%

In addition to the kink waveform of Eq.~(\ref{kink}), 
Eq.~(\ref{dreq2}) gives rise to the
antikink solution, $u_{\mathrm{\,antikink}}(x)=(1/3)\big [1-\tanh (x/3)\big]$. 
As such, for reasons of completeness we also examine kink-antikink interactions. 
The overlap in the kink-antikink pair exponentially decays with the separation
between them, therefore the interaction is appreciable only for a relatively
small separation (i.e., of a few times the width of the individual kink). Actually, this condition is favorable to a potential experimental
realization, as a relatively small spatial domain and short timescales are required. 
An example, when the kink and antikink are brought in close proximity, with
initial separation $\Delta x_{0}=10$, is presented in Fig.~\ref{fig8}(a).
Apparently, the kink-antikink interaction is repulsive, in line with a
similar result known for the cubic-quintic NLS 
model~\cite{filatrella2014domain}; we note here in passing
the similarly repulsive interaction of the (kink-like) dark
solitons of the standard NLS model. 
Specifically, repulsion [Fig.~\ref{fig8}(b)] is accompanied by the spontaneous emission of two counter-propagating dispersive shock waves as can be identified by inspecting the density snapshots shown in Fig.~\ref{fig8}(c). 
We note that the repulsive nature of the interaction persists for moving kink and
antikink, each one initiated by $u(x)=u_{\mathrm{\,(anti)kink}}(x)+\epsilon\,
du_{\mathrm{\,(anti)kink}}(x)/dx$, with boost parameter $\epsilon $ (not shown here). 
Note that, in contrast to the genuinely repulsive OOP droplet-droplet   
interaction [Fig.~\ref{fig4}(b)], here we deal with heteroclinic
waveforms and thus the definition of the associated mass is an 
intriguing problem in its own right; for a recent discussion of the simpler, 
standard GPE problem and its dark soliton heteroclinic solutions, see, e.g., 
Ref.~\cite{pavloff} and, in particular, Appendix C therein. 
Nevertheless, we still observe a repulsive interaction
in Fig.~\ref{fig8}(a), and the accompanying dispersive shock-wave
emission leads the kinks into a slowdown that eventually halts them
in line with the expectation that no traveling such solutions exist
in the model under consideration.

%%%%%%%%%%%%%%%%%%%%%%%%%%%%%%%%%%%%%%%%%%%%%%%%%%%%%%%%%%%%%%%
\section{Conclusions $\&$ perspectives}
\label{conclusions}
%%%%%%%%%%%%%%%%%%%%%%%%%%%%%%%%%%%%%%%%%%%%%%%%%%%%%%%%%%%%%%%

We have studied the static properties of droplets and bubbles  
within the one-dimensional amended GPE model relevant to the inclusion of the LHY term in
homonuclear, mass-balanced, bosonic mixtures.
In this context, we  
systematically explored the interactions among different types of nonlinear excitations. 
Particularly, our investigation involved multiple droplets and bubbles, but also droplet-kink and kink-antikink ones. 
The stability of droplets and bubbles upon chemical potential variations has been explored, expanding upon related earlier work.  
The parametric regions of existence of each solution were identified with droplets transitioning from Gaussian to flat-top ones for decreasing chemical potential while bubbles becoming wider and deeper. 
Moreover, the stability of the droplets and instability of bubbles has been demonstrated by extracting the corresponding Bogoliubov-de-Gennes excitation spectrum. 
The destabilization of bubbles manifests through expansion of their core in the course of the evolution. 

Turning to the dynamics, interactions between droplets were 
found, similarly to bright solitons, to depend on their relative phase and chemical potential difference. 
Importantly, the particle picture for droplets is constructed revealing their in-phase attraction and out-of-phase repulsion.
This outcome is subsequently tested by monitoring the interactions of these entities through the amended GPE model. 
Additionally, we showcase that for intermediate phase differences droplets initially experience attraction followed by repulsion. 
The interaction between droplets with different chemical potentials leads to mass exchange between them and the formation of heavier droplets in an excited state.

On the other hand, bubble interactions, given the bubbles',
individual, unstable nature are highly susceptible to different outcomes 
depending on their initial separation.  
Particularly, if placed in close proximity, they merge into a single bubble due to their mutual attraction followed by the emission of counter-propagating shock waves and the subsequent formation of gray soliton trains. 
However, for slightly increased separation they show individual core expansion trapping in between them a droplet. 
Examining interactions among kinks and droplets reveals the transformation of the latter by the former accompanied by the spontaneous generation of a dispersive shock wave leaving behind a gray soliton. 
These dynamical features are enhanced for traveling or lighter droplets. 
We have also explored kink-antikink interactions which generically result in mutual repulsion and emission of
weak counter-propagating shock waves.

Our findings suggest pathways for further work. A direct extension is to
consider a quasi-1D geometry, aiming to reveal the role of transversal excitations in
colliding droplets and bubbles. 
This is an effort that it appears to us would be particularly useful towards connecting with experimental
settings (including in the presence of external confinement).
Also, a detailed analysis of the emitted
shock-wave structures following the droplet-kink collision is an issue of substantial interest. 
The generalization of the ``two-body" particle-like dynamical picture developed herein for droplets and 
bubbles to 1D {\em chains} of such structures is a promising prospect, producing potentially effective dynamical 
lattice equations similar to the ones for dark and bright NLS solitons described in Ref.~\cite{ric2}.
Other relevant directions are to 
further quantify the analytical
characterization of inter-soliton interactions
(of different type), as well as to
study interactions and collisions between solitons 
and/or vortical droplets in the 2D setting combining the mean-field and LHY terms, and, 
on the other hand, in full quantum states (beyond the limits of the LHY 
description)~\cite{parisi2019liquid,mistakidis2021formation,ota2020beyond}.

%%%%%%%%%%%%%%%%%%%%%%%%%%%%%%%%%%%%%%%%%%%%%%%%%%%%%%%%%%%%%%%
\section*{Acknowledgements}
%%%%%%%%%%%%%%%%%%%%%%%%%%%%%%%%%%%%%%%%%%%%%%%%%%%%%%%%%%%%%%%

S.I.M.~acknowledges support from the NSF through a grant for ITAMP at
Harvard University. This material is based upon work supported by the 
U.S.~National Science Foundation under awards PHY-2110038 (R.C.G.) and
PHY-2110030 and DMS-2204702 (P.G.K.). The work of B.A.M.~was supported, 
in part, by the Israel Science Foundation through grant No.~1695/22.

\bibliographystyle{apsrev4-1}
\let\itshape\upshape
\normalem
\bibliography{reference11}

\end{document}